\documentclass[twocolumn,twoside]{IEEEtran}

\ifCLASSINFOpdf
  
\else
 
\fi{\tiny {\LARGE }}

\ifCLASSOPTIONcompsoc
    \usepackage[caption=false, font=normalsize, labelfont=sf, textfont=sf]{subfig}
\else
    \usepackage[caption=false, font=normalsize]{subfig}
\fi
\usepackage{balance}

\usepackage{multicol}   
\usepackage{cite}
\usepackage{gensymb}
\usepackage{multirow}
\usepackage{graphics}  
\usepackage{epsfig} 
\usepackage{graphicx}
\usepackage{epstopdf}
\usepackage{textcomp}
\usepackage{amsmath}
\usepackage{amssymb}
\usepackage{mathtools,color}
\usepackage{filecontents}
\usepackage{lipsum}

\usepackage{lipsum}%

\usepackage{multicol}   
\usepackage{multirow} 
\usepackage{mathtools}
\usepackage{filecontents}
\usepackage{lipsum,color}
\usepackage{amsfonts}

\usepackage{mathrsfs}
\usepackage{textpos}

\begin{document}
\title{On the Performance of Space-Time MIMO Multiplexing for Free Space Optical Communications}

\author{\IEEEauthorblockN{Mohammad~Taghi~Dabiri,~Himan~Savojbolaghchi~and~Seyed~Mohammad~Sajad~Sadough}\\
	\IEEEauthorblockA{Faculty of Electrical Engineering, Shahid Beheshti University G. C., 1983969411, Tehran, Iran \\
		Email: \{m\_dabiri, h\_savojbolaghchi, s\_sadough\}@sbu.ac.ir}}
\maketitle

\author{Mohammad~Taghi~Dabiri,~Himan~Savojbolaghchi~and~Seyed~Mohammad~Sajad~Sadough 
}

\maketitle
\begin{abstract}
Due to the relatively high cost of deploying optical fibers, free space optical (FSO) links have been developed as a cost-effective alternative technology for next generation cellular networks. However, in order to have a practical role in the physical layer of future communication systems, data rate of FSO links must be improved.
To this aim, in this paper we employ a multiple-input multiple-output (MIMO) multiplexing scheme with two transceivers to increase the data rate of the considered  FSO system.  
Unlike MIMO diversity case, the performance of MIMO multiplexing is significantly affected by interference between parallel channels.
To solve this problem, we propose a novel space-time scheme which significantly reduces the interference between parallel channels. We analyze the performance of the proposed scheme in terms of BER and outage probability. 
\end{abstract}

\IEEEpeerreviewmaketitle

\section{Introduction}
The ever increasing demand for mobile data services and the necessity of more efficient use of the radio spectrum are leading network operators to increase the density of base stations \cite{magazin2015}. This densification has become possible by small-cell deployment. Backhaul is needed to connect the small cells to the core network, internet and other services \cite{magazin2015}.
Optical fiber technologies offer high capacity which are sufficient for next-generation cellular networks.
However, due to its high cost, for some network operators, in some places, the use of fiber is not always affordable. In such scenarios, free space optical (FSO) communication systems have been developed as a cost-effective alternative technology for the backhaul of next-generation cellular networks \cite{jlt2016backhaul,ghassemlooy2015emerging,dabiriJSAC}.
With its significant advantages such as large available bandwidth, low cost implementation, global license-free feature, low risk of exposure and robustness to electromagnetic interference, FSO communication has recently attracted a growing attention for a wide range of applications \cite{khalighi2014,dabiri2017fso}.

To increase the data rate of FSO links, recent efforts show that multiple-input multiple-output (MIMO) spatial multiplexing can significantly improve the transmission rate (see for instance \cite{kahn_2016,khalighi_Max,uysal2016,aghajanzadeh2010, safari2013,huang2017spatial,dabiri2017ergodic,dabiri2017performance}).
In \cite{kahn_2016}, the authors compared the spectral efficiency of conventional MIMO multiplexing and spatial-mode multiplexing with that provided by orbital angular momentum (OAM) multiplexing over turbulence channels.
In \cite{khalighi_Max}, the authors investigated the interest of spatial multiplexing in MIMO FSO systems and compared its performance to those achieved with repetition coding
orthogonal space time block codes (OSTBC) and optical spatial modulation.  
In \cite{aghajanzadeh2010}, the diversity-multiplexing trade-off for log-normal channels were analyzed to optimize both diversity and multiplexing gains when using coherent modulations and heterodyne receivers for FSO systems. 
In \cite{safari2013}, the multiplexing gain has been investigated for MIMO FSO systems when using intensity modulation with direct detection (IM/DD).
More precisely, in \cite{safari2013}, the multiple transmitted beams generate individual airy patterns on the detector plane which are separated due to the difference in the angle of arrival (AOA).
In \cite{huang2017spatial}, spatial-mode multiplexing for practical FSO systems using DD is investigated.
In \cite{dabiri2017ergodic} for two transceiver pairs and in \cite{dabiri2017performance} for $M$ transceiver pairs, the authors have employed a spatial multiplexing scheme to increase the data rate of an FSO system.
More precisely, in \cite{dabiri2017ergodic,dabiri2017performance}, assuming a Gaussian modulation (which is not a practical assumption), the performance of the proposed scheme is analyzed in terms of average achievable data rate.

In order to complete recent results in \cite{dabiri2017ergodic,dabiri2017performance}, in the first part of this paper we analyze the performance of MIMO multiplexing scheme introduced in \cite{dabiri2017ergodic,dabiri2017performance} in terms of BER and outage probability for pulse position modulation (PPM) signaling which is a widely used digital modulation schemes in FSO systems. More precisely, we will show how the tunable parameters such as beam waist at the receiver can affect the performance of the considered system.
Unlike MIMO diversity case, the performance of  MIMO multiplexing is significantly degraded by the interference between parallel channels. To solve this issue, in the second part of this paper, we propose a novel space-time scheme which significantly reduces the interference between parallel channels and improves the performance of MIMO multiplexing case, yet preserving the simplicity of the previous scheme.
\section{System Model}
We consider a MIMO-FSO communication system with two transmitters (two laser sources) and two receiver apertures, where each transmitter sends optical signals toward the center of its corresponding receiver. 
At the transmitters, IM/DD with PPM is exploited to modulate the optical transmitted signals. 
A PPM scheme uses the position of a pulse in two time-slots to represent the value of an information bit, i.e., the presence of a pulse in the first time-slot is characterized by a ``1'' and in the second time-slot is characterized by a ``0''.
The received signal at the $ i $th receiver is denoted $r_i$ for $ i\in \{1 , 2\} $ ($2$ is the number of transceiver pairs) is expressed at any discrete symbol time as
\begin{align}
\label{s4d5}
r_i =& \left[
\begin{array}{rl}  
r_i^{(1)}   \nonumber \\
r_i^{(2)} \end{array} \right]  \nonumber \\
=& \left[
\begin{array}{rl}  
R g_{ii}h_{ii} P_t T_s s_i^{(1)}  + \sum^2_{\substack{j=1 \\ j\neq i}}R g_{ji}h_{ji}P_t T_s s_j^{(1)} + n_i^{(1)}    \\
R g_{ii}h_{ii} P_t T_s s_i^{(2)}  + \sum^2_{\substack{j=1 \\ j\neq i}}R g_{ji}h_{ji}P_t T_s s_j^{(2)} + n_i^{(2)} \end{array} \right].
\end{align}
where $s_i^{(1)}\in\{0,1\}$ and $s_i^{(2)}=1-s_i^{(1)}$ are respectively the transmitted signals in the first and second time-slot corresponding to the BPPM symbol;  $r_i^{(1)}$ and $r_i^{(2)}$ are the received electrical signals in the first and second time-slot.
We consider two spatial signaling method: i) MIMO multiplexing where each transmitter sends independent optical signals and ii) MIMO diversity or repetition coding where all transmitters send same optical signals, i.e., $s_1^{(1)}=s_2^{(1)}$. 
In the case of MIMO diversity, \eqref{s4d5} can be simplified as 
\begin{align}
\label{s4d6}
r_i = \left[
\begin{array}{rl}  
r_i^{(1)}   \\
r_i^{(2)} \end{array} \right]  
= \left[
\begin{array}{rl}  
\sum^2_{j=1}R g_{ji}h_{ji}P_t T_s s_j^{(1)}     + n_i^{(1)}~~~~~~~   \\
\sum^2_{j=1}R g_{ji}h_{ji}P_t T_s (1-s_j^{(1)}) + n_i^{(2)} \end{array} \right],
\end{align}
where $R$ is the photo detector's responsivity, $ P_t $ is the transmitted signal power, $n^{(1)}_i$ and $n^{(2)}_i$ are the signal-independent additive white Gaussian noise (AWGN) with zero mean and variance $ \sigma_{n}^{2} $=$ N_{0}T_s/2 $, $T_s$ denotes the time-slot duration and $h_{ji}$ is the atmospheric turbulence coefficient between $j$th transmitter and $i$th receiver which is assumed perfectly known at the receiver. Notice that this is a practical assumption due to the slow fading property of FSO links  \cite{dabiri2017generalized,dabiri2017glrt,WCL2018}.
Moreover, for $h_{ji}=1$, the fraction of the collected power at $i$th receiver due to transmitted signal by $j$th transmitter can be written as
\begin{align}
\label{Asd4}
g_{ji} =  A_0\exp\left(-2\frac{\left(x_p+d\right)^2+\left(y_p\right)^2}{w_{z_{\rm eq}}^2} \right),
\end{align}
where $[d, 0]$ are distance between two receivers in the $[x,y]$ plane, $ A_0=(\mathrm{erf}(\nu))^2 $ denotes the maximal fraction of the collected power, $ \nu=\frac{\sqrt{\pi}r_a}{\sqrt{2}w_z} $, $ w_{z_{\rm eq}}^2=w_z^2\dfrac{\sqrt{\pi}\mathrm{erf}(\nu)}{2\nu\exp(-\nu^2)} $ is the equivalent beamwidth and $\mathrm{erf}(z)=\frac{2}{\sqrt{\pi}}\int_0^{z}e^{-x^2}dx $ is the error function.
At the receiver aperture plane, we can express the radial displacement vector as $r_p = [x_p, y_p]$, where $x_p$ and $y_p$, denote respectively the displacements located along the horizontal and elevation axes at the receiver which can be modeled as zero mean Gaussian distributed random variables (RV), i.e., $x_p \thicksim  \mathcal{N}(0, \sigma^2_{xp})$ and $y_p \thicksim  \mathcal{N}(0, \sigma^2_{yp})$.

Lastly, we consider the well-known gamma-gamma distribution for modeling the atmospheric turbulence. This way, the PDF of the normalized channel coefficient $h$ is given by \cite{laserbook}
\begin{align}
f_{h}(h)= \frac{2(\alpha\beta)^{\frac{\alpha+\beta}{2}}}{\Gamma(\alpha)\Gamma(\beta)}h^{\frac{\alpha+\beta}{2}-1}
k_{\alpha-\beta}(2\sqrt{\alpha\beta h}), 
\end{align}
where $ \Gamma (.) $ is the gamma function, $ k_{m}(.) $ is the modified Bessel function of second kind of order $ m$, $ 1/\beta$ and $1/\alpha$ are the variances of the small and large scale eddies, respectively. 
\section{Performance Analysis}
%
\subsection{BER Analysis}
%
\subsubsection{MIMO Multiplexing}
According to \eqref{s4d5} and after some mathematical calculations, the average BER of MIMO multiplexing can be obtained as 
\begin{align}
\label{c4v1}
\mathbb{P}_{e,M} = \int_0^\infty \int_0^\infty  \mathbb{P}_{e,M|p}f_{x_p}(x_p) f_{y_p}(y_p) dx_pdy_p,
\end{align}
where
\begin{align}
\label{c4v2}
\mathbb{P}_{e,M|p} = \int_0^\infty \int_0^\infty  \mathbb{P}_{e,M|p,h}f_{h_{11}}(h_{11}) f_{h_{21}}(h_{21}) dh_{11}dh_{21},
\end{align}
and 
\begin{align}
\label{c4v3}
\mathbb{P}_{e,M|p,h}=& \frac{1}{2}Q\left(\frac{R P_t \sqrt{T_s}\left(g_{11}h_{11}+g_{21}h_{21} \right)}{\sqrt{N_0}} \right) \nonumber \\
+&\frac{1}{2}Q\left(\frac{R P_t \sqrt{T_s}\left(g_{11}h_{11}-g_{21}h_{21} \right)}{\sqrt{N_0}} \right).
\end{align}
\subsubsection{MIMO Diversity}
According to \eqref{s4d6}, the BER of MIMO diversity conditioned on $h$ and $r_p$ can be obtained as 
\begin{align}
\label{c4v7}
\mathbb{P}_{e,D|p,h}= Q\left(\frac{R P_t \sqrt{T_s}\sum_{i=1}^2\sum_{j=1}^2 g_{ij}h_{ij}}{\sqrt{2N_0}} \right).
\end{align}
Finally, substituting $\mathbb{P}_{e,D|p,h}$ in \eqref{c4v2} and \eqref{c4v1} instead of $\mathbb{P}_{e,M|p,h}$, the average BER of MIMO diversity is obtained.
\subsection{Outage Probability Analysis}
%
\subsubsection{MIMO Multiplexing}
For a given target BER $\mathbb{P}_{e,t}$, the outage probability is defined as the probability that the communication system can not support $\mathbb{P}_{e,t}$ \cite{dabiri2018Performance}.
According to \eqref{c4v3} and for low values of outage probability (for outage probability lower than $10^{-2}$), the outage probability of MIMO multiplexing case can be closely expressed as
\begin{align}
\label{g2x3}
\mathcal{P}_{\rm out}^M &= {\rm Prob}\left\{ \mathbb{P}_{e,M|p,h}>\mathbb{P}_{e,t} \right\} \nonumber \\
&\simeq  {\rm Prob}\left\{ g_{11}h_{11}-g_{21}h_{21} < \mathcal{A}_{th1} \right\} ,
\end{align}
where $\mathcal{A}_{th1} =\frac{\sqrt{N_0}}{R P_t \sqrt{T_s}}Q^{-1}\left(2 \mathbb{P}_{e,t}\right) $.

\subsubsection{MIMO Diversity}
According to \eqref{c4v7}, the outage probability of MIMO diversity case can be obtained as 
\begin{align}
\label{g2x4}
\mathcal{P}_{\rm out}^D &= {\rm Prob}\left\{ \mathbb{P}_{e,D|p,h}>\mathbb{P}_{e,t} \right\} \nonumber \\
&=  {\rm Prob}\left\{ \sum_{i=1}^2\sum_{j=1}^2 g_{ij}h_{ij} < \mathcal{A}_{th2} \right\} ,
\end{align}
where $ \mathcal{A}_{th2} =\frac{\sqrt{2N_0}}{R P_t \sqrt{T_s}}Q^{-1}\left( \mathbb{P}_{e,t}\right) $.
\subsection{Simulation Results}
In this part, the performance of MIMO multiplexing and MIMO diversity communication systems are numerically studied and the behavior of the considered system is studied versus different tunable parameters such as beam waist at the receiver and distance between receivers.
The values of the parameters used for our numerical analysis are set as follows: aperture radius of receiver $r_a=10$ cm, optical wavelength $\lambda=1.5~{\micro m}$, link length 1 $\rm km$, slot duration $T_s=1$ ns, $\alpha=11.7$ and $\beta=10.2$.

\begin{figure}
	\begin{center}
		\includegraphics[width=3.4 in ]{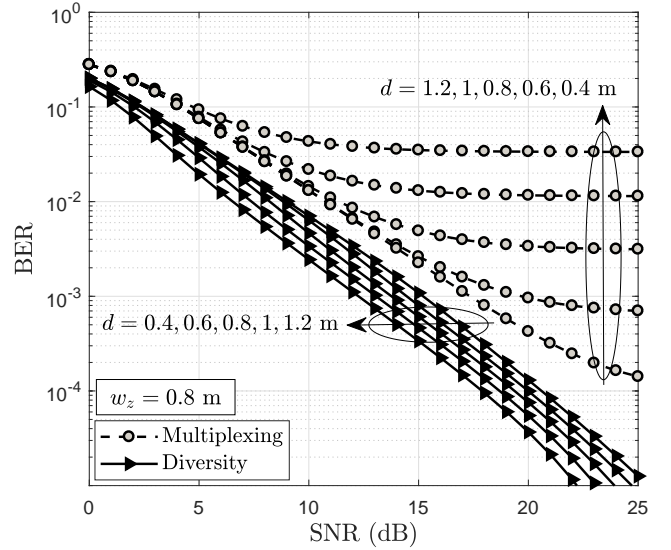}
		\caption{Average BER of MIMO multiplexing and diversity cases versus SNR for different values of  $d$.}
		\label{1}
	\end{center}
\end{figure}

The average BER  of MIMO multiplexing and MIMO  diversity cases versus SNR is depicted in Fig. \ref{1} for different values of $d=1.2, 1, 0.8, 0.6, 0.4~ {\rm m}$.
As expected, MIMO diversity case has better performance compared to the MIMO multiplexing case. However, note that the transmitted rate of multiplexing case is twice as large as diversity case. As we observe from Fig. \ref{1}, by increasing $d$, the performance of MIMO multiplexing case improves, however, the performance of MIMO diversity is a decreasing function of $d$. The reason for this is that the interference between two parallel channels increases by decreasing $d$.

\begin{figure}
	\begin{center}
		\includegraphics[width=3.4 in ]{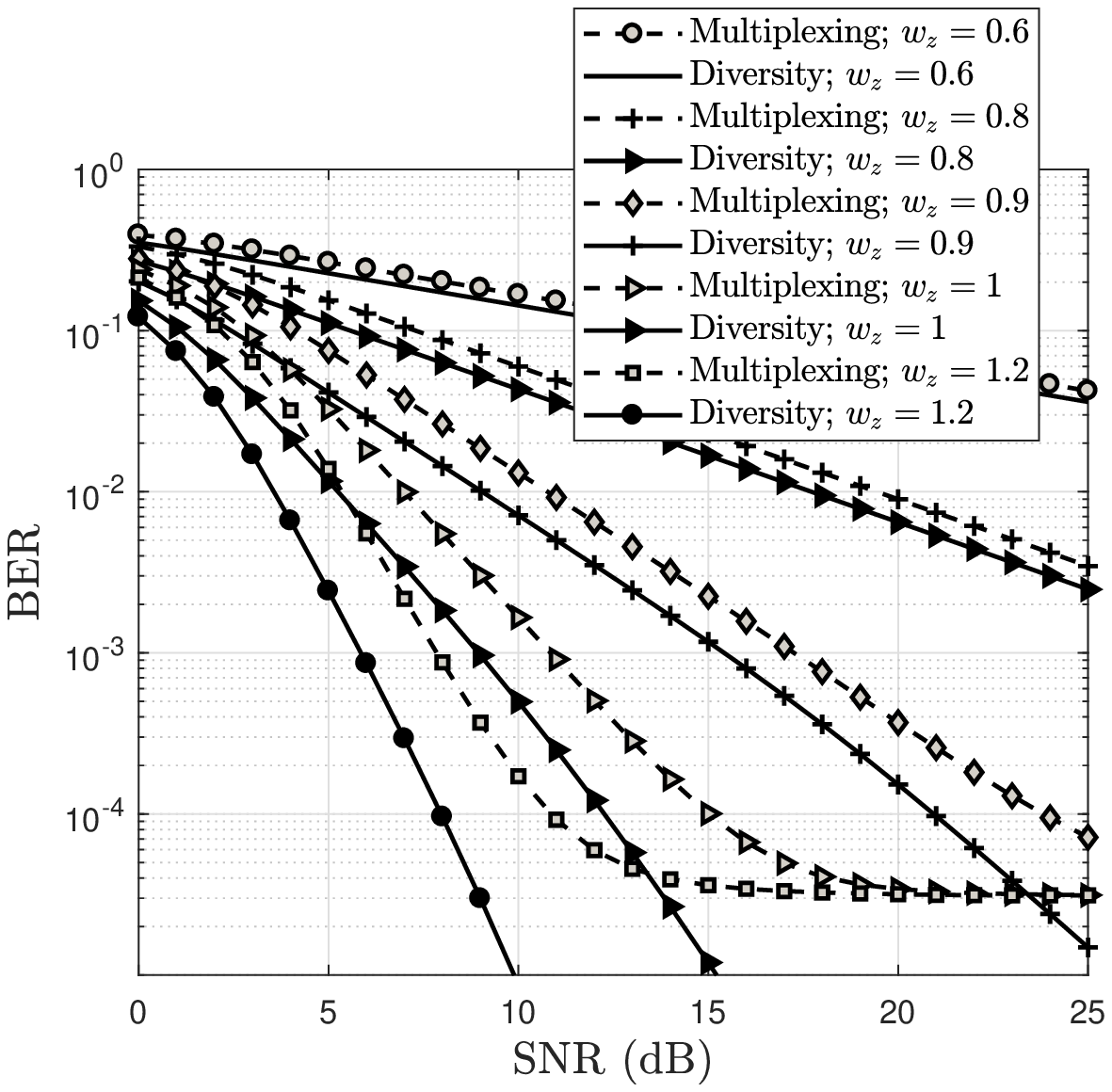}
		\caption{Average BER of MIMO multiplexing and diversity cases versus SNR for different values of  $w_z$.}
		\label{2}
	\end{center}
\end{figure}

\begin{figure}
	\begin{center}
		\includegraphics[width=3.4 in ]{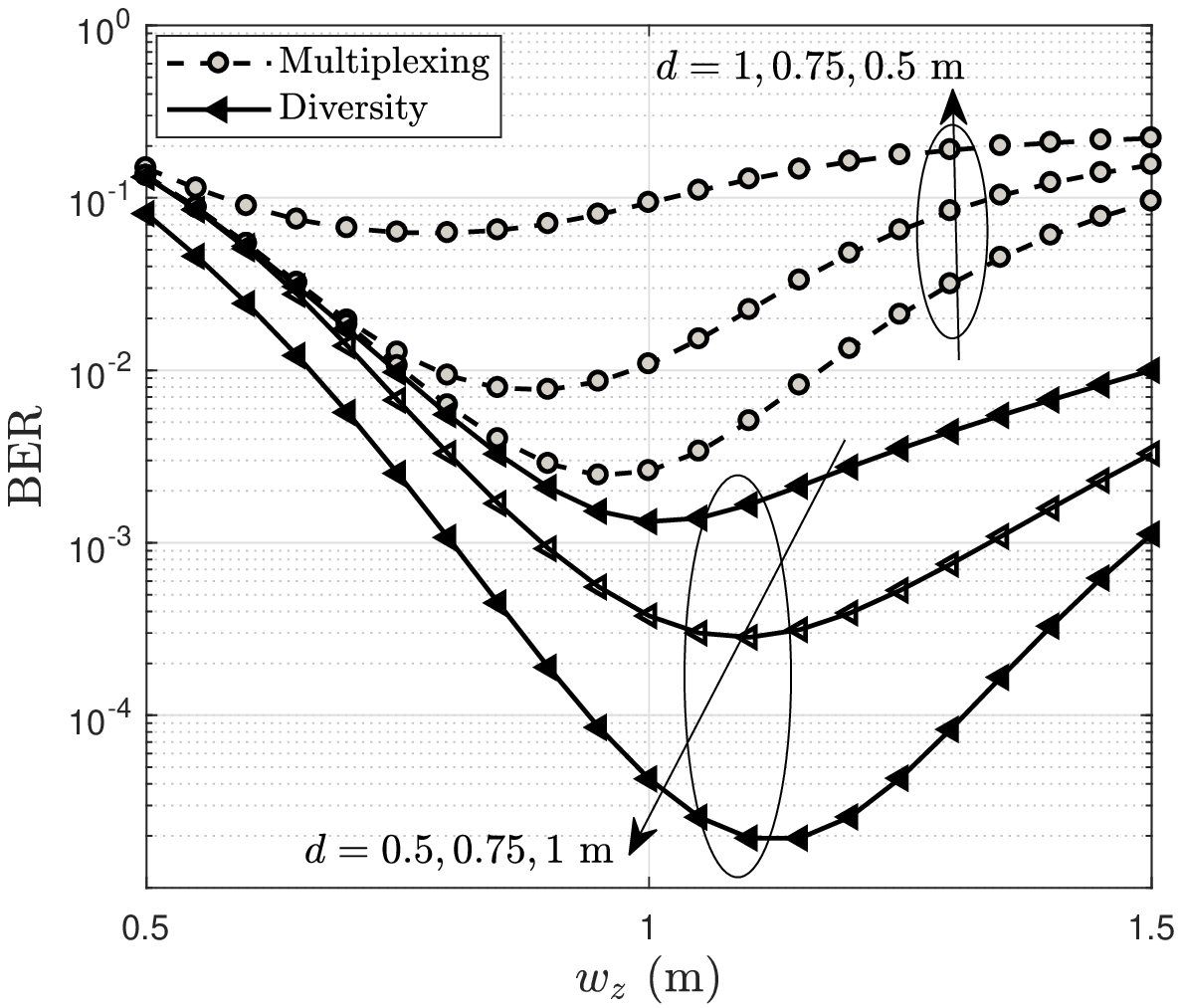}
		\caption{Average BER of MIMO multiplexing and diversity cases versus $w_z$ for different values of  $d$.}
		\label{3}
	\end{center}
\end{figure}

\begin{figure}
	\begin{center}
		\includegraphics[width=3.4 in ]{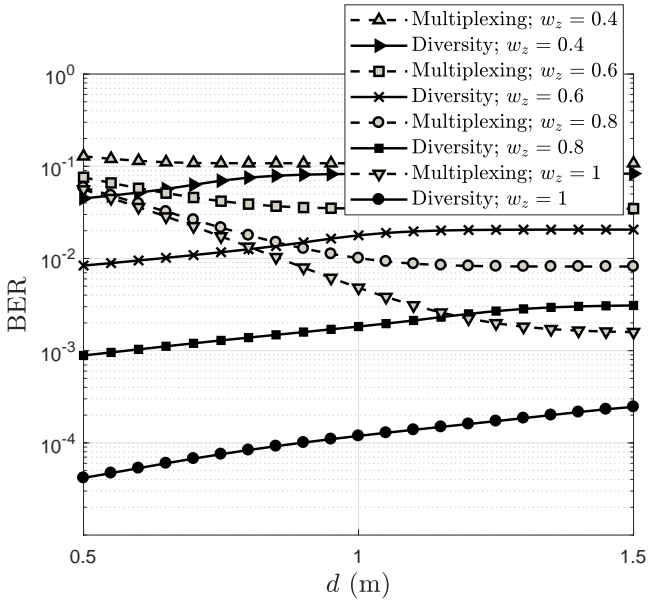}
		\caption{ Average BER of MIMO multiplexing and diversity cases versus $d$ for different values of  $w_z$.}
		\label{4}
	\end{center}
\end{figure}

Another tunable parameter which significantly affects the performance of the considered MIMO system is the optical beam waist at the receiver $w_z$. To show the effect of $w_z$ on the performance of the considered system, the average BER  versus SNR is plotted in Fig. \ref{2} for different values of $w_s=0.6, 0.8, 0.9, 1, 1.2$ m.
As we observe from Fig. \ref{2}, the performance of both multiplexing and diversity system significantly depends on $w_z$. For instance, in the cases of MIMO multiplexing and diversity, the best performance is achieved for $w_z=1.2$. 
At first it might be thought that, the performance of both diversity and multiplexing cases are improved by increasing $w_z$.
To clarify this point, in Figs. \ref{3} and \ref{4}, the BER of considered system are depicted versus $w_z$ and $d$, respectively.
Figure \ref{3} shows that, the optimum value of $w_z$ for both cases depend on $d$. For instance, for MIMO multiplexing case, the optimum values for $w_z$ are 0.95, 0.85 and 0.65 $\rm m$ for $d$=1, 0.75 and 0.5 m, respectively. For MIMO diversity, the optimum values for $w_z$ are 1.15, 1.1 and 1 $\rm m$ for $d$=1, 0.75 and 0.5 m, respectively.
Moreover, results presented in Fig. \ref{4} confirm that the performance of MIMO multiplexing is an increasing function of $d$ while, MIMO diversity is a decreasing function of $d$.
%
%
%
\section{Space-Time Signaling for MIMO Multiplexing}
\begin{figure}
	\begin{center}
		\includegraphics[width=3.4 in ]{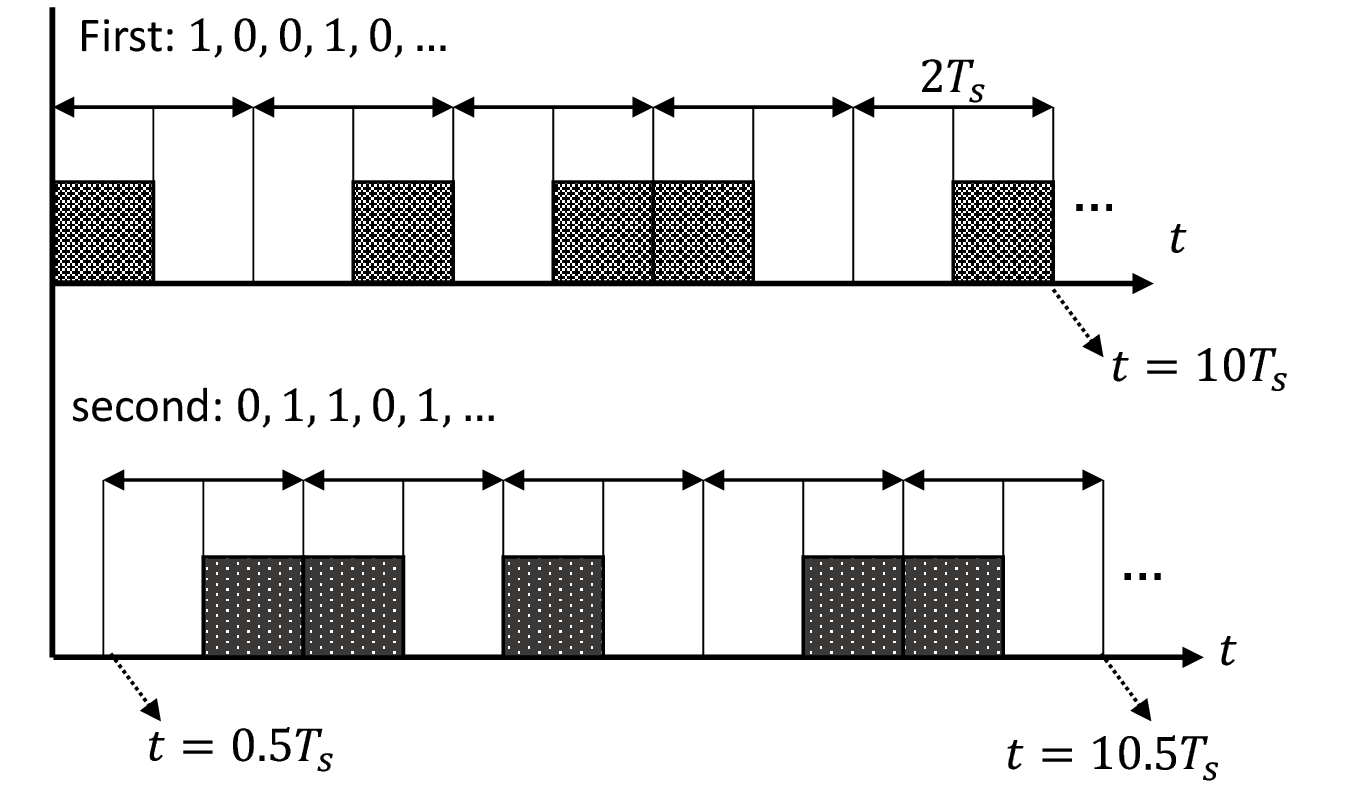}
		\caption{ Start time of first and second transmitters are zero and $T_s/2$, respectively. }
		\label{ds}
	\end{center}
\end{figure}
%
\begin{figure}
	\begin{center}
		\includegraphics[width=3.4 in ]{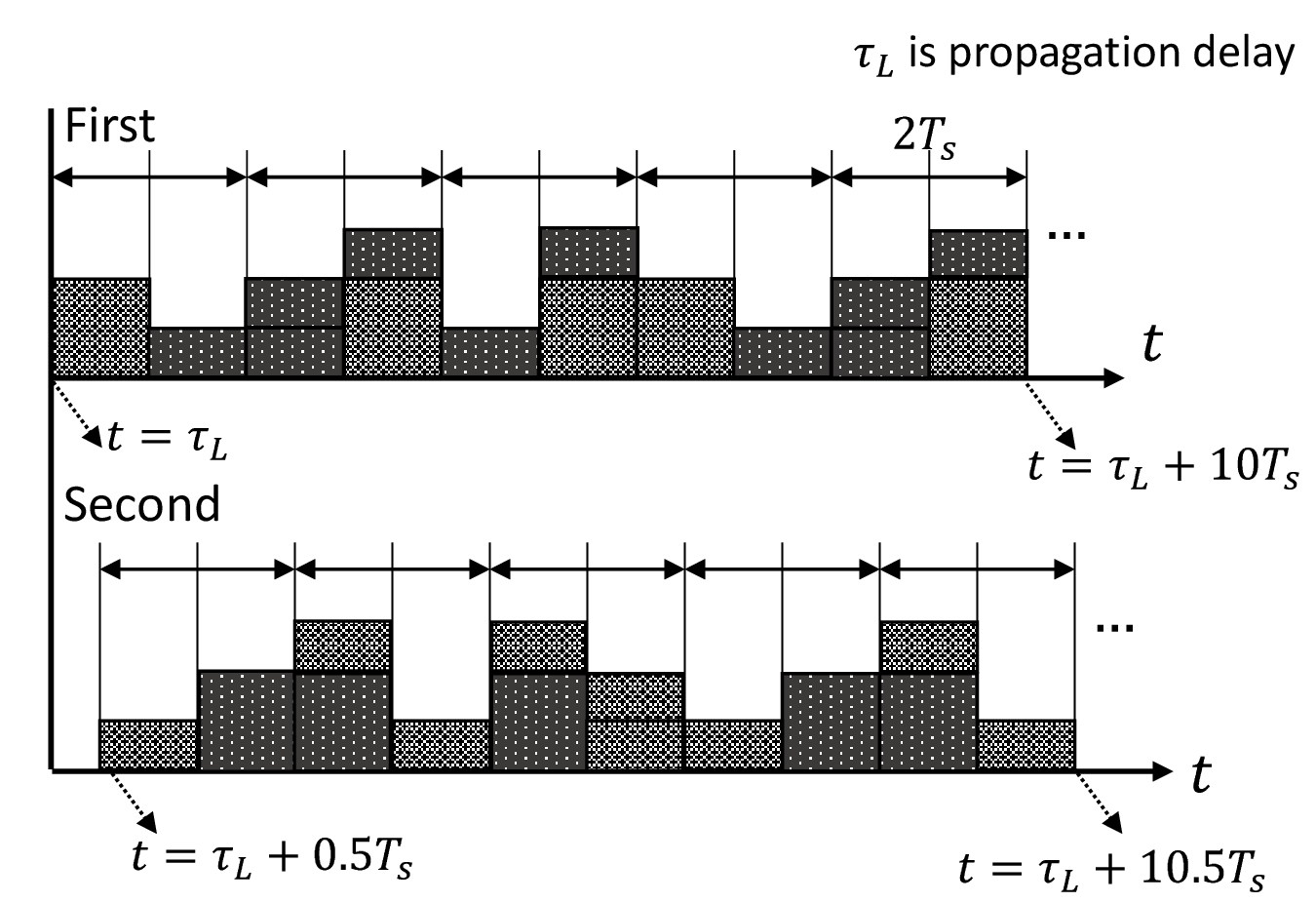}
		\caption{ Received signal equivalent to the transmitted sequences in Fig. \ref{ds} without considering the effect of receiver noises and when $h_{ij}=1$ and $g_{ij}=1$ for $i,j\in\{1, 2\}$. }
		\label{ds2}
	\end{center}
\end{figure}
It is well known that the interference between two parallel channels degrades the performance of MIMO multiplexing case. To reduce the interference between parallel channels, in the sequel we propose a new space-time coding scheme for MIMO multiplexing case when the optical signals are modulated by PPM.

The idea behind this scheme is that the start time of two parallel channels to transfer optical signal are different. We consider the start time of first transmitter is zero and second transmitter is starting to send optical signal after a delay equal to $T_s/2$. Figure \ref{ds} is provided for two independent signal sequence where the first sequence is transmitted by the first transmitter at start time equal to zero and the second sequence is transmitted by the second transmitter at a start time equal to $T_s/2$.  In Fig. \ref{ds2}, we have also depicted the received signal equivalent to the transmitted sequence in Fig. \ref{ds} without considering the effect of receiver noises and when $h_{ij}=1$ and $g_{ij}=1$ for $i,j\in\{1, 2\}$.
This signaling at the transmitter causes approximately same interference in two time-slot of PPM. For instance, when the second transmitter sends bit ``1'', at the first receiver, the interference in first and second slot of PPM are $Rg_{21}h_{21}P_t T_s/2 $. When the second transmitter sends bit ``0'', at the first receiver, the interference depends on the next transmitted bit of second transmitter. For transmitted bits ``0, 0''  and ``0, 1'', the interference at the first slot of first receiver are $ Rg_{21}h_{21}P_t T_s/2$ and $Rg_{21}h_{21}P_t T_s $, respectively, and the interference at the second slot are $ Rg_{21}h_{21}P_t T_s/2$ and $  Rg_{21}h_{21}P_t T_s/2$, respectively.  The BER conditioned on $h$ and $r_p$ of the considered  space-time scheme is obtained as
\begin{align}
\label{fg1s}
\mathbb{P}_{e,ST|p,h}=& \frac{1}{2}Q\left(\frac{R g_{11}h_{11} P_t \sqrt{T_s}}{\sqrt{N_0}} \right) \nonumber \\
&+\frac{1}{4}Q\left(\frac{R g_{11}h_{11} P_t \sqrt{T_s}}{\sqrt{N_0}}  \right) \nonumber \\
&+\frac{1}{8}Q\left(\frac{R P_t \sqrt{T_s}\left(g_{11}h_{11}+g_{21}h_{21}/2 \right)}{\sqrt{N_0}} \right) \nonumber \\
&+\frac{1}{8}Q\left(\frac{R P_t \sqrt{T_s}\left(g_{11}h_{11}-g_{21}h_{21}/2 \right)}{\sqrt{N_0}} \right) \nonumber \\
=&\frac{3}{4}Q\left(\frac{R g_{11}h_{11} P_t \sqrt{T_s}}{\sqrt{N_0}}  \right) \nonumber \\
&+\frac{1}{8}Q\left(\frac{R P_t \sqrt{T_s}\left(g_{11}h_{11}+g_{21}h_{21}/2 \right)}{\sqrt{N_0}} \right) \nonumber \\
&+\frac{1}{8}Q\left(\frac{R P_t \sqrt{T_s}\left(g_{11}h_{11}-g_{21}h_{21}/2 \right)}{\sqrt{N_0}} \right) .
\end{align}  
Finally, substituting $\mathbb{P}_{e,ST|p,h}$ in \eqref{c4v2} and \eqref{c4v1} instead of $\mathbb{P}_{e,M|p,h}$, the average BER of considered space-time scheme is obtained.
For low values of $d$ in which the interference between channels is large, \eqref{fg1s} can be closely approximated as
\begin{align}
\label{g6}
\mathbb{P}_{e,ST|p,h}\simeq \frac{1}{8}Q\left(\frac{R P_t \sqrt{T_s}\left(g_{11}h_{11}-g_{21}h_{21}/2 \right)}{\sqrt{N_0}} \right).
\end{align}

According to \eqref{g6} and for low values of outage probability (for outage probability lower than $10^{-2}$), outage probability of the proposed scheme can be closely obtained as
\begin{align}
\label{g45x3}
\mathcal{P}_{\rm out}^{ST} &= {\rm Prob}\left\{ \mathbb{P}_{e,ST|p,h}>\mathbb{P}_{e,t} \right\} \nonumber \\
&\simeq  {\rm Prob}\left\{ g_{11}h_{11}-g_{21}h_{21}/2 < \mathcal{A}_{th3} \right\} ,
\end{align}
where $\mathcal{A}_{th3} =\frac{\sqrt{N_0}}{R P_t \sqrt{T_s}}Q^{-1}\left(8 \mathbb{P}_{e,t}\right) $.
\subsection{Numerical Results}
\begin{figure}
	\begin{center}
		\includegraphics[width=3.4 in ]{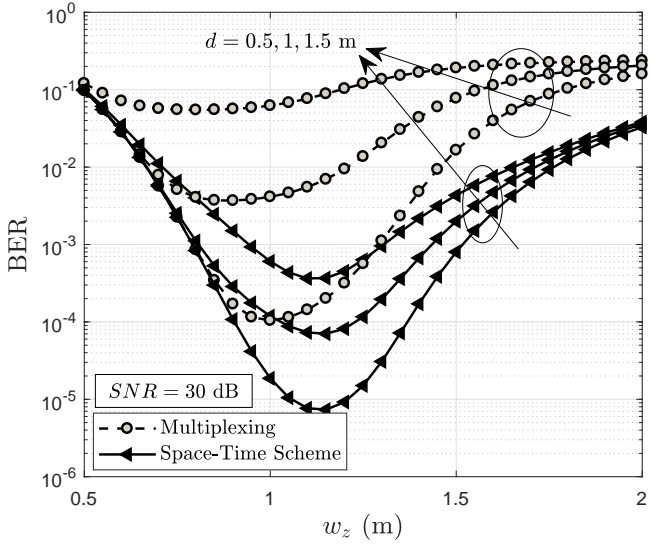}
		\caption{ Comparing average BER of MIMO multiplexing and proposed space-time
			scheme versus $w_z$ for different values of  $d$.}
		\label{5}
	\end{center}
\end{figure}
Assuming here similar system parameters to those introduced in the previous section, in Fig. \ref{5}, the performance of the proposed space-time scheme is contrasted to that of MIMO multiplexing scheme used in conventional PPM. In Fig. \ref{5}, BER obtained with  the considered systems are depicted versus $w_z$ for different values of $d$. Notice that both multiplexing scheme have same bit rate and same complexity and processing load. However, as expected and as we observe from Fig. \ref{5}, by managing the interference between parallel channels, the proposed space-time scheme improves the performance of MIMO multiplexing, significantly.  
\balance 


\end{document}